\documentclass[conference]{IEEEtran}
\IEEEoverridecommandlockouts
\usepackage{cite}
\usepackage{amsmath,amssymb,amsfonts}
\usepackage{algorithmic}
\usepackage{graphicx}
\usepackage{textcomp}
\usepackage{xcolor}
\def\BibTeX{{\rm B\kern-.05em{\sc i\kern-.025em b}\kern-.08em
    T\kern-.1667em\lower.7ex\hbox{E}\kern-.125emX}}

\usepackage{times}
\usepackage{latexsym}
\usepackage{multirow}
\usepackage{times}
\usepackage{latexsym}
\usepackage{multirow}
\usepackage{graphicx}
\usepackage{multirow, boldline}
\usepackage{subfigure}
\usepackage{xcolor}

\usepackage{makecell}
\usepackage{enumitem}

\usepackage{xcolor}
\usepackage{hyperref}

\usepackage{mathtools}

\makeatletter
\newcommand*\concat
  {\mathbin{\mathmakebox[\widthof{${+}\m@th$}]{+\hskip-1emplus1fil+}}}

\makeatother

 \newcommand\citep{\cite}

\usepackage{soulutf8}
\usepackage{todonotes}
\usepackage{placeins}
\usepackage{tikz}
\usepackage{svg}

\title{Low Frame-rate Speech Codec: a Codec Designed for Fast High-quality Speech LLM Training and Inference}
\author{\IEEEauthorblockN{Edresson Casanova, Ryan Langman, Paarth Neekhara, Shehzeen Hussain, Jason Li,\\ Subhankar Ghosh, Ante Jukić, Sang-gil Lee} 
\IEEEauthorblockA{\textit{NVIDIA Corporation}}
}

\begin{document}
%
\maketitle

\begin{abstract}
Large language models (LLMs) have significantly advanced audio processing through audio codecs that convert audio into discrete tokens, enabling the application of language modeling techniques to audio data. However, audio codecs often operate at high frame rates, resulting in slow training and inference, especially for autoregressive models. To address this challenge, we present the Low Frame-rate Speech Codec (LFSC): a neural audio codec that leverages finite scalar quantization and adversarial training with large speech language models to achieve high-quality audio compression with a 1.89 kbps bitrate and 21.5 frames per second. We demonstrate that our novel codec can make the inference of LLM-based text-to-speech models around three times faster while improving intelligibility and producing quality comparable to previous models.
\end{abstract}

\section{Introduction}\label{sec:intro}
Audio codec is an important signal processing technique, that compresses audio signals into discrete codes and then uses these codes to reconstruct the original audio. This technology has long held a central position in fields such as audio transmission and communication \cite{ai2024apcodec}. Recently, it also has been applied to some downstream tasks. For example, some researchers use the discrete codes generated by audio codecs combined with large language models (LLMs), to achieve impressive results in zero-shot text-to-speech (ZS-TTS) \cite{borsos2023audiolm, wang2023neural, zhang2023speechtokenizer, neekhara2024improving} and Speech-to-speech translation (S2ST) \cite{li2023textless, kim2023many, wei2023joint}. 


In recent years, Neural Audio Codecs (NACs) with raw waveform input and output have emerged, offering a balance between decoded audio quality and bitrate \cite{kankanahalli2018end, van2017neural, garbacea2019low, zeghidour2021soundstream, defossez2022high}. For example, SoundStream \cite{zeghidour2021soundstream}, Encodec \cite{defossez2022high}, DAC \cite{kumar2024high},  and  APCodec \cite{ai2024apcodec} use the Residual Vector Quantization (RVQ) \cite{vasuki2006review} to encode audio at low bitrates while using losses from the HiFi-GAN vocoder \cite{kong2020hifi} to maintain audio fidelity. Recent research has focused on improving existing NACs, particularly in the area of quantization strategies. Researchers have worked on improving RVQ, as in \cite{ji2024language} and \cite{luo2024gull}, or exploring new vector quantization approaches. A notable example is the Spectral Codec \cite{langman2024spectral}, which encodes Mel-spectrograms using Finite Scalar Quantization (FSQ). In \cite{neekhara2024improving}, the authors applied a Spectral Codec to the TTS task, demonstrating that it can achieve high-quality audio without relying on techniques like delay patterns \cite{copet2024simple}, which were necessary in previous RVQ-based codecs due to RVQ’s hierarchical architecture. This feature is particularly beneficial for streaming, as it can significantly reduce latency for the prediction of the first audio chunk during inference.  Meanwhile, efforts have been made to introduce or disentangle semantic information during quantization to better suit specific tasks \cite{zhang2023speechtokenizer, ren2024fewer}. Moreover, researchers have worked on improving model structures \cite{xu2024intra}, incorporating additional signal processing techniques in codecs \cite{xiao2023multi} and reducing the computational complexity \cite{xu2024lightcodec}.




{In parallel with our work, researchers have explored the NACs bitrate reduction \cite{ai24b_interspeech, ji2024wavtokenizer}. In \cite{ai24b_interspeech}, the authors integrated APCodec with a bandwidth extension model called AP-BWE\cite{lu2024towards}, which extends bandwidth from 8kHz to 48 kHz. Their proposed model was able to achieve audio coding at 1 kbps bitrate and 25 frames per second (FPS). WavTokenizer \cite{ji2024wavtokenizer} was able to encode audio at 0.9 kbps bitrate using a single codebook at 75 FPS, achieving audio quality comparable with previous SOTA NACs.}

Despite these advancements, some recent works have overlooked the critical importance of low latency, which poses challenges for achieving real-time streamable inference \cite{yang2023hifi, ai2024apcodec, xu2024intra}. For Speech LLMs\footnote{We use the term "Speech LLM" to denote models that integrate LLMs for speech tasks.}, a key factor in achieving low latency is the number of audio frames produced by the codec per second of audio. This is crucial because each frame requires the autoregressive model to perform a forward pass. In scenarios where codes from multiple codebooks can be predicted in parallel, reducing the frame rate becomes even more important than lowering the bit rate. However, most of the current literature focuses primarily on bitrate reduction, often neglecting the potential benefits of frame rate reduction.  For example, the SOTAs codecs DAC and Spectral Codec produce 86 FPS of audio making the inference of Speech LLM models trained with these codecs considerably slow because the Speech LLM needs to do at least 86 forward passes to predict one second of audio. In this paper, we introduce the Low Frame-rate Speech Codec (LFSC) that reduces the frame rate by four times compared to Spectral Codec and DAC, achieving  21.5 FPS making it ideal for Speech LLM model training.

The key contributions of this work are as follows:
\begin{itemize}
 \vspace{-0.05cm}
\item We introduce Low Frame-rate Speech Codec, a novel neural audio codec that compresses audio in  21.5 FPS with a bitrate of {1.89} kbps while maintaining high audio quality.
\item To the best of our knowledge, we are the first to explore large speech language models as discriminators in NACs training, demonstrating significant improvements in low-frame rate scenarios.
\item To demonstrate the effectiveness of our codec in delivering high-quality audio and reducing inference time for Speech LLMs, we trained and evaluated a state-of-the-art (SOTA) LLM-based TTS model, showcasing our codec's performance in practice.
\item Our codec is publicly available in the NeMo repository\footnote{https://github.com/NVIDIA/NeMo}. 
\end{itemize}

The audio samples for each of our experiments are available on the demo website\footnote{https://edresson.github.io/Low-Frame-rate-Speech-Codec}.

\section{Low Frame-rate Speech Codec}\label{sec:model}

Low Frame-rate Speech Codec model is composed of a fully convolutional generator neural network and three discriminators.

The generator comprises an encoder, followed by vector quantization, and a HiFi-GAN-based \cite{kong2020hifi} decoder. The encoder consists of five residual blocks \cite{kong2020hifi}, each block containing three residual layers similar to the multi-receptive field fusion (MRF) module proposed by \cite{kong2020hifi}. However, we use a dilation rate of 1 in the residual layers. To reduce the frame rate to 21.5, each residual block is followed by a 1D convolutional layer with strides of [2, 2, 4, 8, 8] for the respective five blocks. The decoder is based on the HiFi-GAN vocoder with upsampling rates of [8, 8, 4, 2, 2]. The encoder has 48 initial channels, which are doubled after each downsampling layer, while the decoder has 1024 initial channels, which are halved after each upsampling layer. The encoder and decoder have 57.6M and 55.1M parameters, respectively.

For the discriminators, we utilize three neural networks, all employing a squared-GAN and feature-matching loss. We adopt the multi-period discriminator proposed by \cite{kong2020hifi} and the multi-scale complex STFT discriminator proposed by \cite{defossez2022high}. Additionally, inspired by \cite{li2024styletts}, we proposed the use of Speech Language Models (SLMs) as a discriminator. SLMs encode information ranging from acoustic to semantic aspects, which could benefit our model's training, especially in low frame rate settings where accurate pronunciation is difficult to achieve due to the high compression rate. We adopted the 12-layer WavLM \cite{chen2022wavlm}, pre-trained on 94k hours of data, as the SLM. During training, we resample the input audio to 16 kHz before feeding it into the WavLM model, extracting the intermediary layer features. These features are then fed to a discriminative head composed of four 1D convolutional layers. As in \cite{chen2022wavlm}, the SLM remains frozen during training.

For the vector quantization, we followed \cite{langman2024spectral} and used FSQ with eight codebooks and four dimensions per code. However, due to the four times frame rate compression than \cite{langman2024spectral}, we needed to increase the number of codes per codebook from 1000 to 2016\footnote{We have used codebook levels of [8, 7, 6, 6]}, for more details please check Section \ref{sec:ablations}. 

\subsection{Datasets}\label{sec:train}


For codec training, we utilize the same strategy as \cite{langman2024spectral} to get 22.05kHz full-bandwidth audio data. We run bandwidth estimation and apply a bandwidth filter of 11kHz to the English train subset of MLS, and to all languages of Common Voice 13  \cite{ardila2020common}. The Common Voice derived training set comprises 105 languages, totaling 2.7 million utterances, and 3.2k hours of audio from about one-hundred thousand speakers. The MLS English training dataset consists of 6.2 million utterances and 25.5k hours of audio from 4329 speakers.


\subsection{Training setup}\label{sec:train}
We trained our codec in two phases. First, we pre-trained the model with FSQ disabled, and then we fine-tuned it with FSQ enabled. This approach was employed to accelerate the model's convergence when using different quantization techniques. For both pre-training and fine-tuning, the model was trained for approximately 62,000 steps using 96 A100 GPUs with a batch size of 16. The total accumulated batch size was 1,536, and the model processed roughly 95 million samples in each phase. The training was conducted on 1.1-second audio excerpts. We used the Adam optimizer \cite{kingma2014adam} for both the generator and the discriminator, with $\beta_1 = 0.8$, $\beta_2 = 0.99$, and an initial learning rate of 2e-4, which decayed exponentially with a gamma of 0.998.

\subsection{Results and Discussion}\label{sec:results}

We followed an evaluation strategy similar to \cite{langman2024spectral} and employed a combination of objective metrics. For evaluating perceptual quality, we estimate Mean Opinion Scores (MOS) using Torchaudio-Squim \cite{kumar2023torchaudio}. Time-domain accuracy is measured using SI-SDR \cite{le2019sdr}. Spectral accuracy is assessed by calculating the L1 distance between log mel-spectrogram (Mel Dist.) and log magnitude STFT (STFT Dist.) features.
To measure the intelligibility of the codecs reconstruction we compute the character error rate (CER) between the \mbox{Whisper-large v3}\cite{radford2023robust} transcriptions of the ground truth audio and the reconstructed audio.

As the evaluation set, we reconstructed the MLS dataset \cite{pratap2020mls} test set at a 44.1kHz sampling rate by redownloading the audiobooks and filtering out audio files with a bandwidth below 13kHz or a CER exceeding 10\%. The CER was calculated using the \mbox{Whisper-large v3} model. After filtering, we randomly selected 200 samples from each of the eight languages. We chose the MLS test set for its multilingual nature and because we believe that these samples and speakers were not used in the training set of any of the evaluated codecs. This choice is particularly relevant given that most contemporary Speech LLM models are primarily trained on audiobook-like data. The dataset is available on our demo page.

In addition, we assessed the models using the F10 and M10 speakers from the DAPS clear dataset, which had previously been used in the evaluation of the DAC model. We included it to evaluate the models' performance on studio-quality audio. 


{
For evaluation, we selected three SOTA codecs that have been successfully applied in the training of Speech LLM models: Encodec \cite{defossez2022high}, DAC \cite{kumar2024high}, and Spectral Codec \cite{langman2024spectral}.} For Encodec and DAC, we utilized the publicly available checkpoints and we reduced the bitrate via codebook pruning, as done in the original papers. 
For the Spectral Codec, we used the same checkpoint employed in \cite{neekhara2024improving}, as this model has already been evaluated in the context of Speech LLM model training.

Table \ref{tab:code-results} present the results of our evaluation on the 44.1kHz MLS test set and on F10 and M10 DAPS speakers. For codecs operating at a sampling rate lower than 44.1kHz, the audio was resampled to match the target codec's sampling rate. Additionally, during the computation of evaluation metrics, both the ground truth and reconstructed audio were downsampled to 22.05kHz to ensure a more fair comparison.

\begin{table*}
\vspace{-0.2cm}
\caption{Results comparing different codecs using MLS 44.1kHz test set and F10 and M10 DAPS speakers}
\vspace{-0.2cm}
\label{tab:code-results}
\centering
\resizebox{0.98\textwidth}{!}{%
\begin{tabular}{l|c|c|c|c|c|c|c|c|c|c}
\hline
\textbf{Codec}                                                                                     & \textbf{Sampling rate}    & \textbf{Bitrate}                   & \textbf{Token/Sec}            & \textbf{Frames/Sec}            & \textbf{Dataset} & \textbf{Squim MOS ($\uparrow$)} & \textbf{SI-SDR($\uparrow$)} & \textbf{Mel Dist. ($\downarrow$)} & \textbf{STFT Dist.($\downarrow$)} & \textbf{CER ($\downarrow$)} \\ \hline
\multirow{2}{*}{Encodec \cite{defossez2022high}}                                  & \multirow{2}{*}{24kHz}    & \multirow{2}{*}{6kbps}             & \multirow{2}{*}{600}          & \multirow{2}{*}{75}            &       MLS        & 4.15                             & 5.47                        & 0.155                             & 0.070                             &          2.52               \\ \cline{6-11} 
                                                                                                   &                           &                                    &                               &                                & DAPS             & 4.34                             & 6.174                       & 0.168                             & 0.075                             & 0.81                        \\ \hline 
\multirow{2}{*}{DAC \cite{kumar2024high}}                                         & \multirow{2}{*}{44.1kHz}    & \multirow{2}{*}{7.75kbps}          & \multirow{2}{*}{774}          & \multirow{2}{*}{86}            & MLS              & \textbf{4.45}                    & \textbf{10.54}              & \textbf{0.110}                    & \textbf{0.054}                    &       1.57         \\ \cline{6-11} 
                                                                                                   &                           &                                    &                               &                                & DAPS             & 4.65                             & \textbf{12.62}              & \textbf{0.108}                    & \textbf{0.051}                    &        \textbf{0.61}       \\ \hline

\multirow{2}{*}{DAC \cite{kumar2024high}}                                         & \multirow{2}{*}{24kHz}    & \multirow{2}{*}{6kbps}             & \multirow{2}{*}{600}          & \multirow{2}{*}{75}            & MLS              & 4.42                             & 8.19                        & 0.127                             & 0.060                             &              2.26          \\ \cline{6-11} 
                                                                                                   &                           &                                    &                               &                                & DAPS             & 4.57                             & 10.03                       & 0.129                             & 0.059                             &          0.63              \\ \hline
\multirow{2}{*}{DAC \cite{kumar2024high}}                                         & \multirow{2}{*}{24kHz}    & \multirow{2}{*}{3kbps}             & \multirow{2}{*}{300}          & \multirow{2}{*}{75}            & MLS              & 4.30                             & 4.77                        & 0.174                             & 0.071                             &              3.68           \\ \cline{6-11} 
                                                                                                   &                           &                                    &                               &                                & DAPS             & 4.40                             & 5.79                        & 0.182                             & 0.073                             &  0.92                     \\ \hline
\multirow{2}{*}{Spectral Codec \cite{langman2024spectral, neekhara2024improving}} & \multirow{2}{*}{22.05kHz} & \multirow{2}{*}{6.88kbps}          & \multirow{2}{*}{688}          & \multirow{2}{*}{86}            & MLS              & 4.43                             & -20.44                      & 0.117                             & 0.060                             &        \textbf{1.47}        \\ \cline{6-11} 
                                                                                                   &                           &                                    &                               &                                & DAPS             & 4.64                             & -20.43                      & 0.117                             & 0.058                             &         0.67               \\ \hline
\multirow{2}{*}{Ours}                                                                              & \multirow{2}{*}{22.05kHz} & \multirow{2}{*}{\textbf{1.89kbps}} & \multirow{2}{*}{\textbf{172}} & \multirow{2}{*}{\textbf{21.5}} & MLS              & 4.43                             & 4.46                        & 0.147                             & 0.061                             &          2.09              \\ \cline{6-11} 
                                                                                                   &                           &                                    &                               &                                & DAPS             & \textbf{4.68}                    & 6.93                        & 0.142                             & 0.058                             &         0.86               \\ \hline

\end{tabular}
}
\vspace{-0.2cm}
\end{table*}


{Our model achieved the highest perceptual quality score (Squim MOS) on the DAPS test set and the second-highest score on the MLS test set. In terms of SI-SDR, our model outperformed the Spectral Codec across both evaluation sets and exceeded the performance of the 3kbps DAC on the DAPS test set. However, it did not perform as well as the other models. For Mel distance and STFT distance, our model demonstrated better results than both Encodec and the 3kbps DAC, although it performed slightly worse than the other models. Regarding intelligibility (CER), our model surpassed the 3kbps DAC in both sets and outperformed Encodec and the 6kbps DAC in the CML test set.}

These results suggest that our codec is competitive with SOTA codecs, despite its significantly lower bitrate and frame rate. However, strong objective metrics alone do not guarantee superior performance in training Speech LLM models. For instance, \cite{neekhara2024improving} demonstrated that an LLM-based TTS model trained using the Spectral Codec achieved better quality than the same model trained with a 7.75kbps DAC, even though the codec evaluation indicated that the 7.75kbps DAC outperformed the Spectral Codec. In Section \ref{sec:tts-exp}, we applied our codec in the training of a Speech LLM model and highlighted the advantages it offers over other codecs.

The DAC codec's balanced training across various bandwidths renders it robust and capable of producing high-quality output even with low-bandwidth audio. We also evaluated the 7.75kbps DAC model using 22.05kHz audio resampled to 44.1kHz as input (resulting in a bandwidth of approximately 11kHz), on the MLS 44.1kHz test set. When comparing DAC with real 44.1kHz input to the resampled 22.05kHz input, the latter approach yielded identical Squim MOS and Mel distance metrics, with slight improvements in SI-SDR (10.65 vs 10.54) and STFT distance (0.052 vs 0.054), though it exhibited a marginally higher CER (1.62 vs 1.57). This feature is really interesting because it can be used to ensure a fair comparison with the Spectral Codec and our proposed codec, which were trained in 22.05 kHz audio.

 
 



\subsection{Ablation Study}\label{sec:ablations}

We conduct an ablation study of our model, to show the effect of the WavLM-based discriminator and the codebook size of FSQ. During the ablation study, we used the same steps and training approach described in Section \ref{sec:model}. For the training without vector quantization, we have trained the model per 124k steps, which is the sum of the steps from the two training phases. 


\begin{table*}[ht!]
\vspace{-0.2cm}
\caption{Ablation Study results on MLS 44.1kHz test set}
\vspace{-0.2cm}
\label{tab:ablations}
\centering
\begin{tabular}{l|l|l|l|l|l}
\hline
 \textbf{Condition} &  \textbf{Bitrate} & \textbf{Squim MOS} & \textbf{SI-SDR} & \textbf{STFT Dist.}  & \textbf{CER } \\ \hline
         Our model                  &    \textbf{1.89kbps}     &      4.43  &    4.45               &  0.061            &      2.09 \\ \hline
         Our model  W/o vector quantization   &  -                 &   \textbf{4.45}   &   \textbf{7.06}    &  \textbf{0.058}    &   \textbf{1.73}              \\ \hline
 Our model  W/o WavLM disc.   & \textbf{1.89kbps}                  &  4.42    &      {5.04} & {0.060}      &             2.46      \\ \hline
      Our model  FSQ 1000 codes       &   \textbf{1.72kbps}        &   4.43  &    4.10     &             0.062              &    2.61   \\ \hline
       Our model  FSQ 4032 codes       &     \textbf{2.06kbps}        &   \textbf{4.45}  &      4.86    &    0.061                      &    {1.96}   \\ \hline
\end{tabular}
 \vspace{-0.45cm}
\end{table*}

Table \ref{tab:ablations} presents our ablations study results.  The WavLM-based discriminator slightly improved the general quality (Squim MOS 4.43 vs 4.42). It also enhanced speech intelligibility with a good margin (CER 2.09 vs 2.46). However, it leads to a slightly worse time-domain accuracy (SI-SDR 5.04 vs 4.45) and STFT distance (0.061 vs 0.060).

{The increase in codebook size from 1000 codes to 2016 produced the same score for general audio quality. However, it improved the SI-SDR, STFT distance, and especially speech intelligibility (CER 2.61 vs 2.09). Increasing the codebook size even more from 2016 to 4032 brings slightly better results for almost all the metrics. In addition, using FSQ with  4032 codes our model was able to achieve the same general quality as the model trained without vector quantization and competitive performance in the other metrics.}


\section{ZS-TTS study}\label{sec:tts-exp}

\subsection{Experiments setup}\label{sec:tts-exps-setup}

To evaluate how our codec performs in comparison to previous codecs in practice, we replicated the experiments outlined in \cite{neekhara2024improving}. We utilized the T5-TTS \cite{neekhara2024improving} model with context conditioning on the decoder due to its superior performance on the ZS-TTS task.

For training the TTS model, we employed the same datasets as those used in \cite{neekhara2024improving}. These datasets comprise 1.8k hours of English data from four sources: the train-clean 360 subset of LibriTTS~\cite{zen2019libritts}, HiFiTTS~\cite{bakhturina21_interspeech}, a 1000-hour subset from MLS train set~\cite{pratap2020mls}, and a proprietary dataset featuring two speakers and totaling 63 hours. 


{The model was trained using three SOTA codecs: Encodec, DAC, Spectral Codec, and two variants of our proposed codec. Although models such as APCodec and WavTokenizer have recently demonstrated promising performance, they have not yet been explored in the context of Speech LLM training. Considering that these codecs may require specific tricks to work properly in this context, we decided to focus on SOTA codecs that have been successfully applied in the training of Speech LLM models.}

For training, we utilize a fixed context duration of three seconds, where context is an alternate utterance from the speaker of the target utterance.  We train each of our models with a batch size of 192 distributed across 32 NVIDIA A100 GPUs, for 300,000 steps. The training is conducted using the AdamW optimizer with a fixed learning rate of 1e-4.  During inference, we use multinomial Top-k sampling with k equal to 80 and temperature equal to 0.85. 

Following~\cite{neekhara2024improving} and prior works, for the 24kHz 6kbps Encodec and 44.1kHz 7.75kbps DAC codecs, we utilized the delay pattern method~\cite{copet2024simple} for the codec sequence. For both the Spectral Codec and our proposed codec, we predicted all eight codes in parallel. To ensure a fair comparison with the Spectral Codec and our proposed codec, and in accordance with~\cite{neekhara2024improving}, we extracted the codes from 22.05kHz audio resampled to 44.1kHz for the DAC model.

\subsection{Results and Discussion}\label{sec:tts-results}

To evaluate the performance of the T5-TTS model with various codecs, we followed the methodology proposed by~\cite{neekhara2024improving}.{We consider 200 utterances from the VCTK dataset~\cite{veaux2017cstr}.}
These sentences were generated using 20 random unseen speakers with 10 utterances per speaker from the VCTK dataset. Consistent with~\cite{neekhara2024improving}, we used Character Error Rate (CER), Speaker Encoder Cosine Similarity (SECS), and Mean Opinion Score (MOS) as evaluation metrics.

CER was computed by comparing the transcription of the generated audio with the TTS input text, utilizing the Conformer-Transducer Large ASR model\footnote{https://huggingface.co/nvidia/stt\_en\_conformer\_transducer\_large}. For SECS, we computed the cosine similarity between the speaker embeddings of the synthesized speech and the target speaker reference provided to the TTS model, using the WavLM speaker verification model\footnote{https://huggingface.co/microsoft/wavlm-base-plus-sv} to extract the speaker embeddings.

For the MOS evaluation, 200 audio samples were generated for each model. Each listener was presented with one sample and asked to rate it on a scale from 1 to 5, in 1-point intervals. Each audio sample was evaluated by at least 10 independent listeners.

In addition to these metrics, we computed the inference real-time factor (RTF). For better readability, we normalized the RTF values by setting the fastest model's RTF to 1 and adjusting the other models' RTF values accordingly. 

Table \ref{tab:results-tts} presents the results of our ZS-TTS experiments.

\begin{table}[ht!]
\vspace{-0.2cm}
\caption{T5-TTS evaluation using different codecs}
\vspace{-0.2cm}
\label{tab:results-tts}
\centering
\resizebox{0.48\textwidth}{!}{%
\begin{tabular}{l|l|l|l|l|l}
\hline
 \textbf{Codec} &  \textbf{Bitrate} &  \textbf{MOS($\uparrow$)} & \textbf{CER($\downarrow$)} & \textbf{SECS($\uparrow$)} & \textbf{RTF($\downarrow$})  \\ \hline
Encodec   &   6kbps      &   3.52 $\pm$ 0.04         &  2.97     &   0.749   &  2.839  \\ \hline
 DAC   & 7.75kbps   &  3.94 $\pm$ 0.04   &    2.66    &  0.709   &  3.318 \\ \hline
       Spectral  Codec    &     6.88kbps    &        3.87 $\pm$ 0.04           &       2.23             & 0.794   &  3.255 \\ \hline 
      Our Codec     &   1.89kbps    &    3.95 $\pm$ 0.04        &   \textbf{0.93}                &   \textbf{0.823} &  \textbf{1.000} \\ \hline
    \begin{tabular}[c]{@{}l@{}}Our Codec -\\ 4032 FSQ\end{tabular}    &   2.06kbps   &  \textbf{4.02 $\pm$ 0.04}   &  1.19          &        0.797              &  1.003  \\\hline
\multicolumn{6}{c}{\footnotesize * Note that  RTF values are normalized  by setting the fastest model’s RTF to 1.}
\end{tabular}
}
\vspace{-0.35cm}
\end{table}

The T5-TTS model trained using our proposed codec achieved better CER, showing that our codec improved the pronunciation accuracy in the TTS task. Our proposed codecs achieved performance comparable to previous models in terms of MOS. No significant differences were observed when comparing our codecs with DAC and Spectral Codec. In addition, the model achieved better speaker similarity, while bringing around 3 times speedup compared to other SOTA codecs. 

Although, in Section \ref{sec:ablations}, our proposed codec using 4032 codebook size achieved better metrics than the proposed codec with 2016 codebook size, this improvement in performance did not generalize for the TTS downstream task. Despite, the model 4032 codebook size achieving a better MOS, there is no significant difference. In addition, the model with a smaller codebook size produced better performance in all the others metrics. We believe that it happens because when the codebook size is smaller the model tends to generalize better. A similar pattern was also reported in \cite{casanova2024xtts}.

\section{Conclusions and future work} \label{sec:conc}

In this work, we presented the Low Frame-rate Speech Codec, which achieved high-quality audio compression at a bitrate of {1.89} kbps and 21.5 frames per second. Furthermore, we demonstrated that a Speech LLM model trained with our codec achieved quality comparable to that of the same model trained with previous state-of-the-art audio codecs, while providing approximately threefold speedup. 

In future work, we intend to explore our approach in the 44 kHz audio codec training and also investigate the application of our codec to other audio domains, such as music and sound effects.

\vfill\pagebreak



\bibliographystyle{IEEEtran}
\bibliography{refs}


\end{document}